\documentclass[12pt]{article}
\usepackage[koi8-r]{inputenc}
\usepackage[saorus]{sao1}
\usepackage{graphicx}

\begin{document}
\begin{center}
{\Large \bf Stellar Subsystems of the Galaxy NGC~2366.}\\
\bigskip

{\large N. A. Tikhonov$^1$ and   O. A. Galazutdinova$^1$}

\bigskip

{\em $^1$ Special Astrophysical Observatory, Russian Academy of Sciences,
Nizhnii Arkhyz, 357147, Karachai-Cherkessian Republic, Russia}
\end{center}

{AstronomyReports, 2008}

\begin{abstract}
Hubble Space Telescope archive data are used to perform photometry of stars in seven
fields at the center and periphery of the galaxy NGC~2366. The variation of the number
density of stars of various ages with galactocentric radius and along the minor axis of the
galaxy are determined. The boundaries of the thin and thick disks of the galaxy are found.
The inferred sizes of the subsystems of NGC~2366 ($Z_{thin} = 4$ kpc and $Z_{thick} = 8 $ kpc for
the thin and thick disks, respectively) are more typical for spiral galaxies. Evidence for a
stellar halo is found at the periphery of NGC~2366 beyond the thick disk of the galaxy.
 \end{abstract}
 -----------------------------  \\
 $^*$Contact e-mail: dolly@sao.ru
\begin{center}
{\large \bf INTRODUCTION}
\end{center}

\bigskip
\noindent

   Studies of the stellar populations in irregular galaxies have shown that young stars, and, to a
lesser extent, old stars, are concentrated toward the center, with old stars extending to much
greater galactocentric distances than blue stars [1---6]. The number densities of old stars in
individual galaxies decrease exponentially with galactocentric radius [6---9]. Based on an
analysis of the space distribution of stars of various ages in nine spiral and 24 irregular galaxies,
we proposed empirical models of the structure of the distribution of stars in spiral and irregular
galaxies [9, 10], which reflect the characteristics of the distributions of stars of various ages.
The resulting models for the distributions of stars in spiral and irregular galaxies are similar,
and differ mostly in the spatial sizes of stellar subsystems and the presence a stellar halo
in massive spiral galaxies.

   Intermediate-type galaxies (between irregulars and spirals) fell beyond the scope of our
previous studies. Such galaxies are most often represented by massive irregulars or late-type
dwarf spirals with inconspicuous arms. However, studies of intermediate-type galaxies are important
in connection with the question of why spiral, but not irregular, galaxies have halos. Studies of
individual galaxies suggest that there must be some relation between the mass of a galaxy and
the presence or absence of a halo at its periphery. Some massive irregulars (IC~10, M~82) have
conspicuous halos [9, 11], whereas no halos have been found in some dwarf spirals. Revealing
the origins of halos requires increasing the number of transition-type galaxies studied.

   NGC~2366 is of special interest among intermediate-type stellar systems. A fairly large number
of images taken in various fields across the periphery of the galaxy are available from the Hubble
Space Telescope (HST) archive (Fig.~1). NGC~2366 is usually classified as an irregular (NED),
although Baade [12] considered it to be a low-luminosity spiral galaxy. NGC~2366 belongs to the
M81 group (Table ~1), and is mostly interesting because of the giant star-forming region NGC~2363
at its periphery. The stellar population of the central regions has been studied in detail by a number
of authors [14---17], but no studies of the stellar populations at the galaxy periphery have been made previously.

   The $UBVJHK$ and $H_{\alpha}$ images of NGC~2366 show a stellar disk with a well-defined
oval shape [18], which is viewed at an angle of nearly $90^{\circ}$, whose brightness decreases
exponentially from the center toward the edge. The observed asymmetric distribution of young stars
along the major axis of the galaxy is due to star-forming regions. Star formation is especially violent
in the region of the giant $H~II$ complex NGC~2363, but, on the whole, the star-formation rate in
NGC~2366 does not appreciably exceed the average rates in other galaxies of the same type [18].

   The outer structures of the galaxy are conspicuous in the $HI$ map, where we can see two
 extended branches aligned parallel to the major axis of the galaxy [18]. The behavior of the azimuthally
averaged $HI$ surface density in NGC~2366 resembles the behavior of the $HI$ surface density in
other irregular galaxies, except for the outer parts of NGC 2366, where the $HI$ density decreases
more slowly and emitting regions extend farther from the center [18]. The outer boundary of the $HI$
distribution agrees approximately with the outer parts of the optical image, but is asymmetric with
 respect to the $V$ isophote. Richer and Sancisi [19] point out that this is a common feature in disk galaxies.

\vspace{0.1cm}

\begin{center}
{\large \bf STELLAR PHOTOMETRY}
\end{center}

 We consider here images adopted from the HST archive (Table~2), and used the
MIDAS  DAOPHOT~II [20] and HSTphot [21, 22] programs to perform stellar photometry. 
We calculated the completeness of the sample of stars identified using the standard method
of artificial stars, adding several hundred artificial stars of various luminosities to the sample. 
The accuracy of the photomeric zero point in HSTphot is about $0.05^m$  [22], and this 
determines the photometry accuracy for all bright stars. We use the recommendations of 
Holtzmann et al. [23, 24] when transforming the instrumental DAOPHOT~II magnitudes
 into the standard Kron-Cousins $VI$ system.
   
   DAOPHOT~II and HSTphot photometry yield virtually identical results for bright stars in HST
images. However, HSTphot photometry is more accurate internally for faint stars, as is evident
from the decrease in the width of the red-giant branch obtained using the different programs
applied to stars in the same field. This increase in the internal accuracy is most likely due to the
use of stricter star selection criteria in the final photomeric list, and the more refined technique
used to estimate the sky background in the vicinity of the stars measured. However, DAOPHOT~II
photometry has deeper limiting magnitude and can be used for images taken in a single filter,
 when HSTphot cannot be applied.

\vspace{0.1cm}
\begin{center}
{\large \bf  RESULTS OF STELLAR PHOTOMETRY}
\end{center}

   Figure 2 shows the results of our photometry for stars in NGC~2366 in the form of color---magnitude
 (CM) diagrams. The diagrams for the central regions of the galaxy (fields SI and S2) do not differ
 from those of starburst galaxies. We can see branches populated by (1) young stars --- blue and red super-giants;
 (2) intermediate-age stars --- the asymptotic giant branch (AGB), and (3) old stars --- the red-giant branch (RGB).
The CM diagrams for fields located at large galactocentric distances (S3, S4, S6, S7) show only RGB stars, with
few AGB stars.

The distance of NGC~2366 has been determined using various methods (Table~3).
The populated red-giant branch, which is conspicuous in the CM-diagrams (Fig.~2), can be used to find the
distance via the Tip of the Red-Giant Branch (TRGB) method [28], which was used to by Thuan et al. [26]
and Karachentsev et al. [27]. Both teams based their estimates on only a single field, whereas we have
used images of four fields. The tips of the red-giant branches of fields SI, S2, S3, and S4 were determined
by applying a Sobel filter to the corresponding luminosity functions (see [29]). We took the halfwidth of
the Sobel filter to be the uncertainty of the tip of the red-giant branch. Abrupt changes in the slopes of
the luminosity functions of fields SI, S2, S3, and S4 in Fig. 3 corresponding to the tip of the red-giant
branch are visible at $I = 23.56^m, 23.50^m, 23.55^m,$ and $23.55^m$. The average value is
$I_{TRGB} = 23.54^m\pm0.04^m$. The extinction toward NGC 2366 is $A_V = 0.12^m [13]$.
We used the equations of Lee et al. [28] to infer the distance modulus $m - M = 27.48^m \pm 0.17^m$,
corresponding to a distance of $D = 3.13 \pm0.25$ Mpc. The Final distance uncertainty includes
the uncertainties in the TRGB method ($0.10^m$), the TRGB magnitude ($0.04^m$), and the HSTphot
method ($0.05^m$). The distance error also includes the error of the red-giant photometry, since this
 leads to some scatter of the stars in the CM diagram and broadens the Sobel function. The average
 accuracy of the WFPC2 (HST) photometry of a single star is $0.06^m$ at $I = 23.5^m$, but the accuracy of our
 result is increased by our use of more than a dozen stars at the red-giant boundary to calculate the tip
 of the red-giant branch.

 Our distance estimate for NGC~2366 agrees with that of Karachentsev et al. [27], but not with the value
 obtained by Thuan and Izotov [26]. This discrepancy must be due to the use of different calibration methods.
 Our photometry yields low metallicities for red giants, $[Fe/H] = -1.96$, consistent with the low metallicity
of the galaxy determined earlier using spectroscopic methods.

\vspace{0.1cm}
 \begin{center}
{\large \bf DISTRIBUTION OF STELLAR NUMBER DENSITY AND THE SIZESOF THE STELLAR SUBSYSTEMS.}
\end{center}

   We chose two directions to study the visible distributions of stars across the body of the galaxy:
 perpendicular to the equatorial plane of the galaxy (along the Z axis) and in the radial direction, from
 the center of the galaxy outward. Our choice of this latter, not very convenient, direction is due to the
 orientation of the images taken with the HST telescope.

   We used our CM diagrams to select stars of various ages. Figure 2 (field S2) shows regions occupied
  by young stars --- blue supergiants (BSG); intermediate-age stars  --- AGB stars; and old stars --- red giants (RGB).
 We identify these domains in the CM diagrams for each field and use the stars falling within these domains to
 determine their number densities along the galactocentric radius and the Z axis. The only exception is field S7,
 for which deeper-exposure images are available. In this field, we were able to use fainter stars than in other fields
 of the galaxy to analyze the stellar number-density distributions.

   Note that the AGB and RGB stars cannot be unambiguously distinguished at the upper boundary of the RGB,
 since the AGB also passes near the RGB, making it impossible to distinguish the two branches. The AGB stars
 affect the results only slightly, since they are scarse and their number does not exceed 10\% of the total number
  of red giants. We therefore ignored this effect.

  The boundary of the distribution of young stars, i.e., the boundary of the thin disk of NGC 2366, can easily be
  determined by analyzing the results for fields S3 and S4, which lie at this boundary. The number of blue stars
 in Figs. 4c and 4d decreases to zero at distances from the galactic plane of $Z = 1.5$ and 2.2 kpc, respectively.
  The AGB stars in these fields extend somewhat farther than the blue stars, but are less
numerous, and fluctuations in their number density result in greater uncertainty in the inferred boundary of
their distribution. The gradient of the number density of red giants in these fields is small, indicating the large
 extent of the old stellar subsystem.

   A region with a high density of young stars can be seen in Fig. 4a at a galactocentric distance of 0.9 kpc against
the overall decrease in the young-star number density with galactocentric distance. This region coincides with one
of two hydrogen bridges extending parallel to the major axis of the galaxy. Hunter et al. [18] and Thuan et al.[30]
interpret the extended hydrogen structures as an $HI$ ring tilted 60$^\circ$ to the line of sight. The overdensity
of blue stars we have found in the region of this hypothetical ring (Fig. 4a) allows us to view the situation from
a different angle. First, this extended stellar and gaseous structure may be associated with tidal interaction between
NGC~2366 and its neighboring galaxies (e.g., with the nearby and more massive spiral galaxy NGC~2403). Second,
 it is possible that we do not view NGC~2366 at an angle of 90$^\circ$, and that the extended stellar and gasesous
 formations noted above are simply the spiral arms of NGC~2366. Other such galaxies with inconspicuous spiral
 structures are found in the M~81 group (NGC~4236 and IC~2574).

   Unlike the number density of young stars, the number density of red giants does not increase in the region of the
hypothetical ring or spiral arms (Fig.~4a). This is consistent with the hypothesis that the galaxy may have a spiral
structure, since the spiral galaxy NGC 300 shows a similar red-giant distribution un-correlated with the locations of
spiral arms [31].

   Field S5, which is located at a greater galactocentric distance, has been imaged only in a single filter. We nevertheless
analyzed the distribution of stars in this field. Since this region may contain only foreground stars and red giants of
NGC~2366, we obtained a list of stars consisting mostly of red giants by selecting on the luminosity function for the
S5 stars the luminosity interval between the boundary of the tip of the red-giant branch and the photomeric limit with
50\% sample completeness (based on photometry of artificial stars). The distribution of the number density of stars
selected in this way shows a monotonic density decrease with radius (Fig.~4f), without any obvious change in the gradient.
The very small extrapolation of the distribution of stars of this field to zero number density indicates the nearness of
the thick-disk boundary.

   We drew the final boundaries between the stellar subsystems (thin and thick disks) based on our analysis of the
distributions of the stars in all the fields studied. We proceeded from the assumption that the stellar disks are symmetric.
Images of new fields can alter the locations of the boundaries only slightly. This conclusion is corroborated by a recently
found single image taken with the ACS/WFC camera of the HST telescope in the $R$ filter near field S7. This image
confirms the correctness of the thick-disk boundary drawn in this paper.

   The distribution of red giants with galactocentric radius in the field S7, located far from the center of the
   galaxy (Fig.~4j), shows a well-defined change in the number-density gradient at a galactocentric distance of 6.1 kpc.
 We conclude, by analogy with the distributions of stars in spiral and irregular galaxies [6, 9, 31], that the observed break
in the distribution of the red-giant number density coincides with the boundary between the thick disk and halo. We
verified this hypothesis by constructing the distribution of stellar number density in this field perpendicular to the radial
direction (Fig.~4i). The distribution of stars in this direction shows no gradient, as we would expect if the stars
 in field S7 form a disk structure (thick disk and halo) that is symmetric about the center of NGC~2366.

  The distribution of red giants in field S6, which is the farthest from the center (Figs.~4g, 4h), shows no obvious
 radial variations of the stellar number density. This seems to indicate either a lack of stars of NGC~2366 in this field
or very small gradients of the decrease of the number density of halo stars. In any case, to improve the results,
photometry of stars in larger areas of the galactic halo and longer exposures are required to increase the number 
of halo stars involved.

\begin{center}
{\large \bf  FOREGROUND STARS.}
\end{center}

   The effect of foreground stars on our results is significant only for the small statistical sample of red giants we
have studied. This effect should show up when analyzing stars in the halo, which has a low stellar density. To
approximately estimate the effect of foreground stars, we analyzed HST images of areas with Galactic latitudes
 close to that of NGC~2366 [32]. The results suggest that the number of foreground stars is insignificant, and that
these stars form no clumps in the CM diagrams. At the same time, despite their spatially tenuous distribution and
the presence of significant photometric errors due to their faintness, stars in the galaxy periphery do form clumps in
the CM diagrams that are concentrated toward the RGB.

Since NGC~2366 is located fairly far from the equatorial plane of our own Galaxy ($b= 28.5^\circ$ for NGC~2366), 
the effect of foreground stars should be negligible. An analysis of the CM diagram of the field S6, which is farthest
from the galactic center, shows that, even if they are present, RGB stars in this field are very sparse, despite the
long exposures used to take the images in this field (Table~2). We can therefore treat, with appropriate caution,
our results for this field as the results for a foreground field in the close vicinity of NGC~2366.

 \begin{center}
{\large \bf RESULTS AND DISCUSSION.}
\end{center}

   (1) Our stellar photometry in several fields of the galaxy NGC~2366 has enabled us to refine the distance to 
   this galaxy. Our new distance estimate, $D = 3.13 \pm0.25$  Mpc, agrees with that of Karachentsev et al. [27], 
   but differs significantly from estimate of Thuan and Izotov[26].

   (2) Our analysis of the visible distributions of stars of various ages has revealed the thin disk, thick disk, and halo
 stellar subsystems in NGC~2366.

   (3) We have determined the sizes of the thin and thick disks along several directions from the center of the galaxy,
 as well as the boundaries of these subsystems assuming symmetry of the stellar disks (Fig.~1). We find the
 thicknesses of the thin and thick disks in the $Z$ direction to be 4 and 8 kpc, respectively.

   (4) Based on our results for the distribution of young stars, we interpret the $HI$ bridges visible in the galaxy $HI$ 
   image as weak spiral arms, and suggest that NGC~2366 is a low-mass spiral galaxy.

   (5) Our results for the distribution of red giants in the field S7 suggest the existence of a halo in NGC~2366. 
   However, further observations of larger fields with deeper photometric limits are needed to confirm this conclusion,
    and to determine the size of the halo.

  The sizes of the stellar subsystems in NGC~2366 that we have obtained are not consistent with the average
 sizes for subsystems in irregular galaxies [6, 9], but agree better with the sizes of subsystems in spiral galaxies
 [31], lending indirect support for the hypothesis that NGC~2366 is a spiral galaxy. We showed earlier, based on
 similarities of their stellar structures, that subdividing galaxies into spirals and irregulars can be fairly arbitrary, 
 and that both classes of galaxies are more correctly called disk galaxies. Our analysis of the distribution of stars
 in NGC~2366 demonstrates again an absence of differences in morphology between the stellar structure of
 spiral and irregular galaxies.

\begin{center}
{\large \bf ACKNOWLEDGMENTS.}
\end{center}

   This work was supported by the Russian Foundation for Basic Research (project code 03-02-16344).
   This research has made use of information from the NED database (NASA/IPAC Extragalactic Database).

\newpage
\begin{center}
{\large \bf REFERENCES}
\end{center}

\bigskip

1.  D. Minniti and A. Zijlstra, Astron. J. 114, 147(1997).

2. D. Martinez---Delgado, K. Gallart, and A. Aparicio, Astron. J. 118,862(1999).

3. D. Minniti, A. Zijlstra, and V. Alonso, Astron. J. 117, 881 (1999).

4. A. Aparicio and N. Tikhonov, Astron. J. 119, 2183 (2000).

5. Y. Momany, E. V. Held, I. Saviane, and L. Rizzi, Astron. Astrophys. 384, 393 (2002).

6. N. A. Tikhonov, Astron. Rep. 49, 501 (2005).

7. R. Lynds, E. Tolstoy, E. J. O'Neil, and D. Hunter, Astron. J. 116, 146(1998).

8. V. Vansevicins, N. Arimoto, T. Hasegava, et al., Astrophys. J. 611, L93 (2004).

9.  N. A. Tikhonov, Astron. Rep. 50, 517(2006).

10. N. A. Tikhonov and O. A. Galazutdinova, Astrofizika 48, 261 (2005).

11. I. O. Drozdovsky, R. E. Schulte-Ladbeck, U. Hopp, et al., Astron. J. 124, 811 (2002).

12. W. Baade and C. H. Payne-Gaposhchkin, Evolution of Stars and Galaxies (Cambridge, Harvard University, 1963).

13. D. J. Schlegel, D. P. Finkbeiner, and M. Davis, Astrophys. J. 500, 525 (1998).

14. A. Sandage and G. A. Tammann, Astrophys. J. 191, 603 (1974).

15. N. A. Tikhonov, B. I. Bilkina, I. D. Karachentsev, and T. B. Georgiev, Astron. Astrophys., Suppl. Ser. 89, 1 (1991).

16. A. Aparicio, J. Cepa, C. Gallart, et al., Astron. J. 110, 212 (1995).

17. E. Tolstoy, A. Saha, J. G. Hoessel, and K. McQuade, Astron. J. 110, 1640 (1995).

18. D. A. Hunter, B. G. Elmegreen, H. van Woerden, Astrophys.J. 556, 773 (2001).

19. O.-G. Richerand, R. Sancisi, Astron. Astrophys. 290, L9 (1994).

20. P.  B.  Stetson,  Users Manual for DAOPHOT~II (1994).

21. A. E. Dolphin, Publ. Astron. Soc. Pac.  112. 1383 (2000).

22. A. E. Dolphin, Publ. Astron. Soc. Pac.  112, 1397 (2000).

23. J. A. Holtzmann, J. J. Hester, and S. Casertano, Publ. Astron. Soc. Pac. 107, 156 (1995).

24. J. A. Holtzmann, C. J. Burrows, S. Casertano et al., Publ. Astron. Soc. Pac. 107, 1065 (1995).

25. G.de Vaucouleurs, Astrophys. J. 224, 710 (1978).

26. T. X. Thuan and Y. I. Izotov, Astrophys. J. 627, 739 (2005).

27. I. D. Karachentsev, A. E. Dolphin, D. Geisler, et al., Astron. Astrophys. 383, 125 (2002).

28. M. G. Lee, W. L. Freedman, and B. F. Madore, Astron. J. 417, 553 (1993).

29. B. F. Madore and W. L. Freedman, Astron. J. 109, 1645 (1995).

30. T. X. Thuan. J. E. Hibbard, and F. Levrier, Astron. J. 128, 617 (2004).

31. N. A. Tikhonov, O. A. Galazutdinova, and I. O. Drozdovsky, Astron. Astrophys. 431, 127 (2005).

32. 0. A. Galazutdinova, PhD Theisis, Spec. Astrophys. Obs., Russ. Acad. Sci., Nizhnii Arkhyz, (2005).
\newpage
\begin{table}
\begin{center}
\footnotesize
\renewcommand{\tabcolsep}{4pt}
\caption{Basic data for NGC~2366}
\vspace{0.2cm}
 \begin{tabular}{|crrrlrrrrrr|} \hline \hline
\multicolumn{1}{|c}{R.A.(2000.0)}&
\multicolumn{1}{c}{DEC.(2000.0)}&
\multicolumn{1}{c}{$V_h$}&
\multicolumn{1}{c}{$B_t^0$}&
\multicolumn{1}{c}{Type}&
\multicolumn{1}{c}{$A_v$}&
\multicolumn{1}{c}{$A_i$}&
\multicolumn{1}{c}{$i$}&
\multicolumn{1}{c}{$m-M$}&
\multicolumn{1}{c}{$M_{B}$} &
\multicolumn{1}{c|}{}\\ \hline
 $07^h28^m54.^s66$ &$+69^{\circ}12^{\prime}56.^{\prime\prime}8$ &100 &10.95&IB(s)m & 0.120& 0.070& 90.0& 27.48& $-$16.53& \\
\hline
\multicolumn{10}{l}
{ The $V_h$ and $B_t^0$ magnitudes and thr galaxy type are adopted from the NED database.}\\
\multicolumn{10}{l}
{ The extinction coefficients are adopted from [13].}\\
\multicolumn{10}{l}
{ The inclination of the galaxy $i$ is adopted from the LEDA database.}\\
\multicolumn{10}{l}
{The distance modulus $(m-M)$ and absolute magnitude $M_{B}$ are determited in this paper.}\\
\multicolumn{10}{l}
{}\\
\end{tabular}
\end{center}
\end{table}

\begin{table}
\begin{center}
\caption{Log of archive data of WFPC2(HST).}
\vspace{0.2cm}
\renewcommand{\tabcolsep}{3pt}
\begin{tabular}{|ccccccr|} \hline \hline
\multicolumn{1}{|c}{}&
\multicolumn{1}{c}{Date of}&
\multicolumn{1}{c}{}&
\multicolumn{1}{c}{Galactocentric}&
\multicolumn{1}{c}{Exposure}&
\multicolumn{1}{c}{Number of }&
\multicolumn{1}{c|}{Number of}\\
\multicolumn{1}{|c}{Region}&
\multicolumn{1}{c}{observation}&
\multicolumn{1}{c}{Filter}&
\multicolumn{1}{c}{ distance}&
\multicolumn{1}{c}{ }&
\multicolumn{1}{c}{HST}&
\multicolumn{1}{c|}{stars on the}\\
\multicolumn{1}{|c}{}&
\multicolumn{1}{c}{}&
\multicolumn{1}{c}{}&
\multicolumn{1}{c}{(arcmin)}&
\multicolumn{1}{c}{}&
\multicolumn{1}{c}{proposal}&
\multicolumn{1}{c|}{CM diagram}\\ \hline

 S1   & 2000-09-18 & F814w&  0.72  &  600  & 8601& 11403\\
              & 2000-09-18 & F606w&  0.72  &  600  & 8601&\\
        S2   & 2000-12-12 & F814w&  1.20  &  4100  &8769& 17165\\
             & 2000-12-11 & F555w&  1.20  &  6700  &8769&\\
        S3   & 2002-01-08 & F814w&  3.42  &  2$\times$1000   &9318& 6008\\
             & 2002-01-08 & F606w&  3.42  &  2$\times$700    &9318& \\
        S4   & 2001-12-29 & F814w&  3.66  &  2$\times$1000   &9318& 2853\\
             & 2001-12-29 & F450w&  3.66  &  2$\times$1000   &9318&\\
        S5   & 1999-07-27 & F606w&  7.00  &  2$\times$500    &8090& $-$\\
             & 1999-07-27 & F606w&  7.00  &  2$\times$1000   &8090&\\
             & 1999-07-27 & F606w&  7.00  &  2$\times$1200   &8090&\\
             & 1999-07-27 & F606w&  7.00  &  2$\times$1500   &8090&\\
        S6   & 1996-04-19 & F814w&  10.03 &  3400            &5971& 450\\
             & 1996-04-19 & F606w&  10.03 &  7900            &5971& \\
        S7   & 1996-12-07 & F814w& 6.54   &  15000           &6802& 1648\\
             & 1996-12-07 & F606w& 6.54   &  4200            &6802&\\

\hline
\end{tabular}
\end{center}
\end{table}

\begin{table}
\begin{center}
\caption{ Published   distance-modulus   estimates   for NGC 2366.}
\vspace{0.2cm}
\begin{tabular}{|ccc|} \hline \hline
\multicolumn{1}{|c}{Distance modulus}&
\multicolumn{1}{c}{Method}&
\multicolumn{1}{c|}{Reference}\\ \hline

 27.07 & blue supergiants    &  [13]  \\
 27.10 & blue supergiants    &  [32]  \\
 27.62 & blue supergiants    &  [14]  \\
 27.68 & Cepheids              &  [16]  \\
 27.67 & red supergiants     &  [26]  \\
 27.52 & red supergiants     & [25]   \\
\hline
\multicolumn{3}{l}
{Our distance modulus estimate is $(m-M) = 27.48 $}\\
\end{tabular}
\end{center}
\end{table}

\newpage
\begin{figure}[h]%
\centerline{\includegraphics[width=13cm, bb=37 10 572 664,clip]{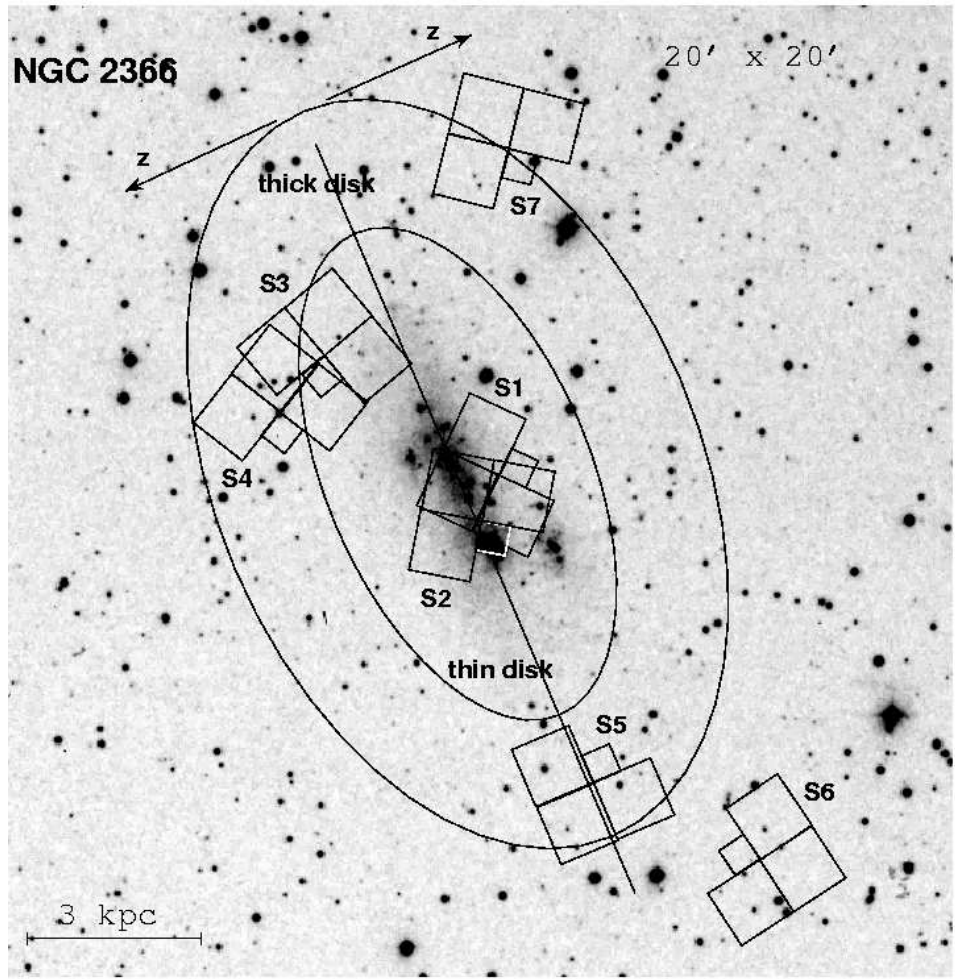}}%
\caption{ Image of NGC 2366 adopted from the DSS-2 survey. The studied areas,
imaged with HST/WFPC2, and the directions Z along which the distribution of stellar number
density was analyzed are shown. The inner ellipse delimits the thin-disk region. The outer
 ellipse corresponds to the boundary between the thick disk and halo.}
\end{figure}

\begin{figure}[h]
\centerline{\includegraphics[angle=270, width=15cm, bb=86 52 567 620,clip]{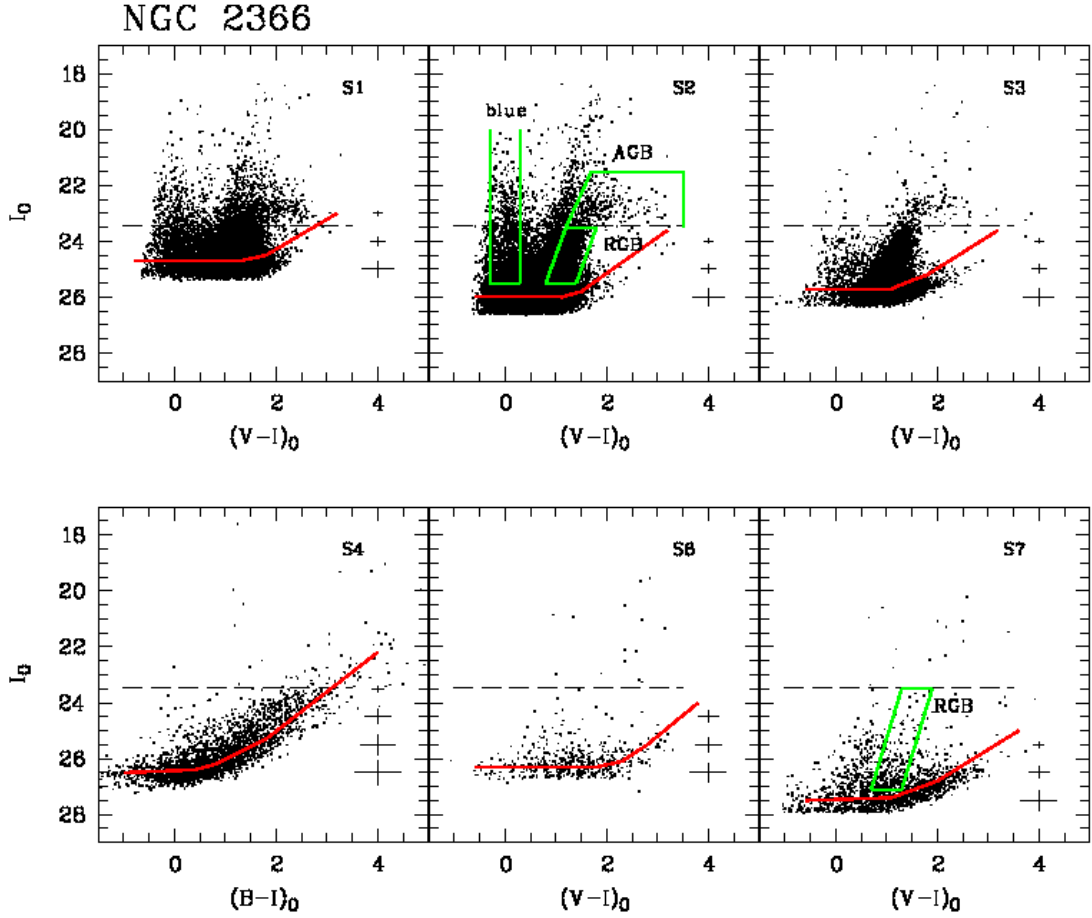}}%
\caption{Color-magnitude diagrams for the fields studied (SI, S2, S3, S4, S6, S7) in
NGC 2366. The dashed lines show the extinction-corrected position of $I_{TRGB}^0 = 23.^m47$
and the solid line the 50\% completeness level for the stellar sample determined using an
artificial-stars test. The photometric error boxes are shown. The domains of stars of various
ages are shown in diagrams for areas S2 and S7: young blue stars (blue), intermediate-age stars (AGB),
and old stars (RGB).}
\end{figure}

\begin{figure}[h]
\centerline{\includegraphics[angle=0, width=14cm, bb=64 411 472 851,clip]{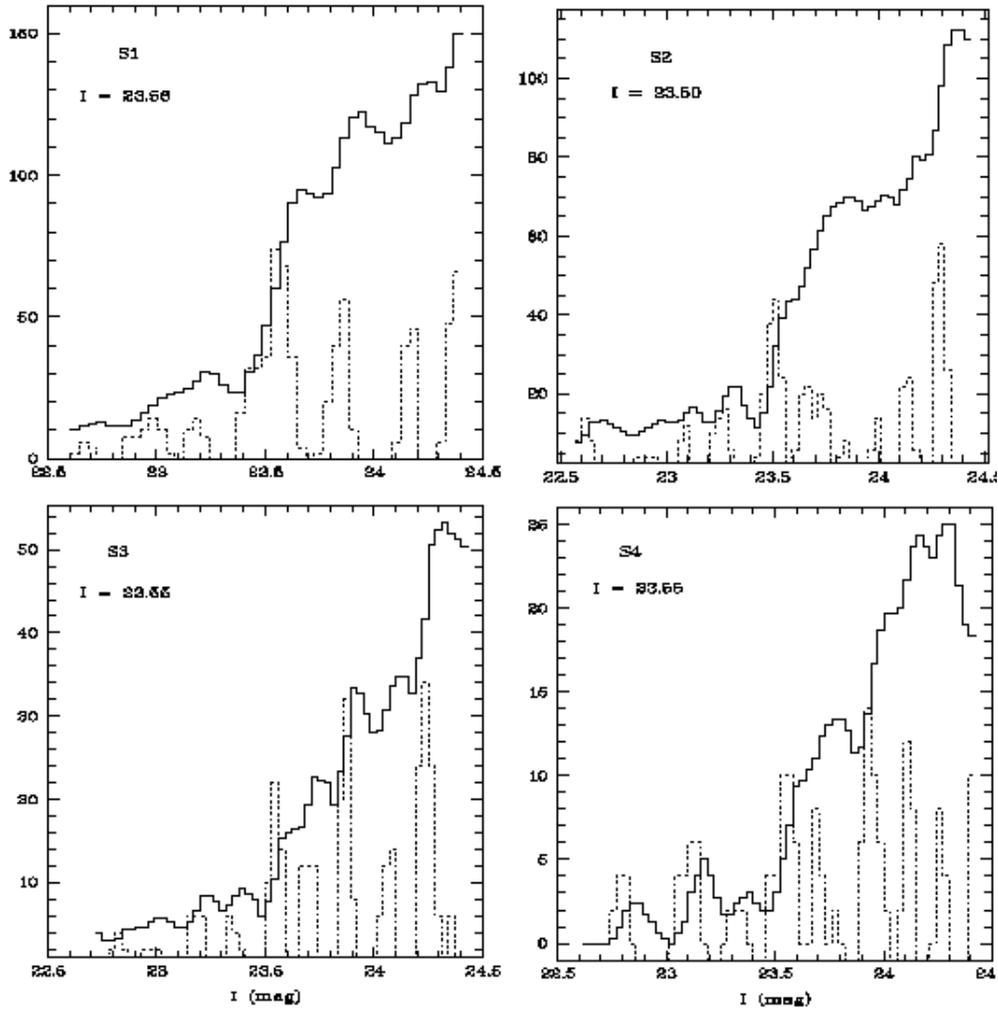}}
\caption{$I$-band luminosity functions for $(V - I > 1.0)$ stars in the fields studied (SI, S2, S3, S4).
The abrupt change in the number of stars corresponds to the start of the red-giant branch,
which is used in the TRGB distance-determination method [28].}

\end{figure}
\begin{figure}[h]
\centerline{\includegraphics[angle=0, width=15cm, bb=25 5 590 820,clip]{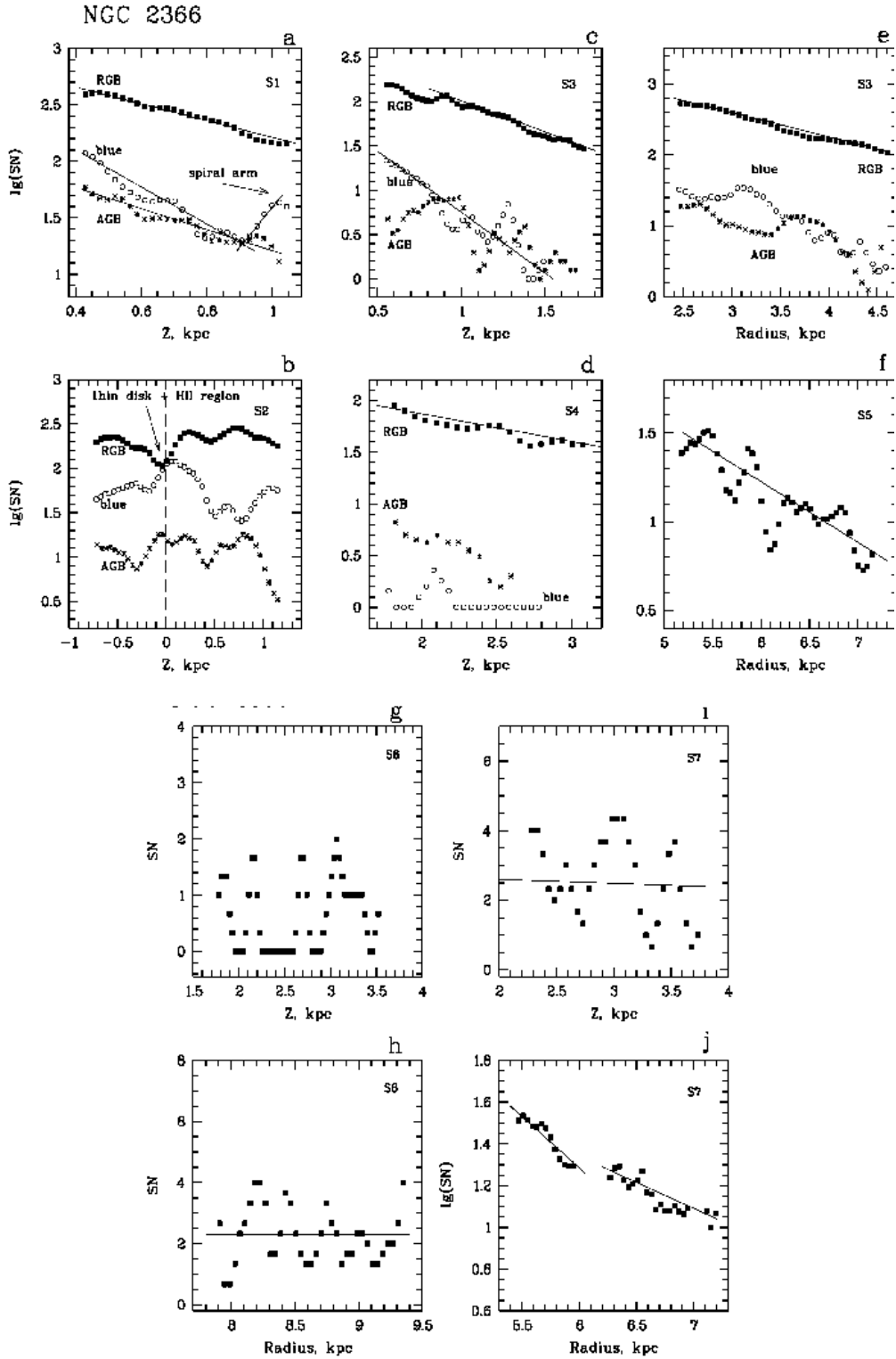}}%
 \caption{Distribution of the number density N of stars of various types in the NGC~2366 fields
 studied (S1, S2, S3, S4, S5, S6, S7) along the $Z$ axis and the radial direction $R$. The Filled squares,
 open circles, and asterisks show red giants (RGB), blue stars, and AGB stars, respectively.}
\end{figure}

\end{document}